\documentclass[epj,twocolumns]{svjour_mod}

\usepackage[utf8]{inputenc}
\usepackage{latexsym}
\usepackage{graphics}
\usepackage{amssymb}
\usepackage[colorlinks,citecolor=blue,urlcolor=blue,linkcolor=blue]{hyperref}
\usepackage{hyphenat}
\usepackage{amsmath}
\usepackage{cite}
\usepackage{relsize}
\usepackage{trimclip}
\usepackage{booktabs} 
\usepackage{tabularx}
\usepackage{multirow}
\usepackage{fixmath}
\usepackage{tikz}
\usetikzlibrary{calc}
\usepackage{placeins}
\usepackage{xspace}
\usepackage{slashed}
\newcommand{\be}{\begin{equation} \begin{aligned}}
\newcommand{\ee}{\end{aligned} \end{equation}}
\newcommand{\beqa}{\begin{eqnarray}}
\newcommand{\eeqa}{\end{eqnarray}}

\newcommand{\mev}{\text{MeV}}
\newcommand{\gev}{\text{GeV}}

\newcommand{\iab}{\text{ab}^{-1}}

\newcommand{\ifb}{\text{fb}^{-1}}

\newcommand{\cm}{\text{cm}}
\newcommand{\m}{\text{m}}

\RequirePackage[normalem]{ulem}

\begin{document}


\title{{\Large Predictions for Neutrinos and New Physics from Forward Heavy Hadron Production at the LHC}}

\author{Luca Buonocore$^1$
\and Felix Kling$^2$
\and Luca Rottoli$^1$
\and Jonas Sominka$^3$}

\institute{Physik Institut, Universit\"at Z\"urich, CH-8057 Z\"urich, Switzerland
\and
Deutsches Elektronen-Synchrotron DESY, Notkestr.~85, 22607 Hamburg, Germany
\and
Physik Institut, Universit\"at Z\"urich, CH-8057 Z\"urich, Switzerland
\and 
Institute of Theoretical Physics, Universit\"at Hamburg, 22761, Hamburg, Germany}

\abstract{
 Scenarios with new physics particles feebly interacting with the Standard Model sector provide compelling candidates for dark matter searches. Geared with a set of new experiments for the detection of neutrinos and long-lived particles the Large Hadron Collider (LHC) has joined the hunt for these elusive states. On the theoretical side, this emerging physics program requires reliable estimates of the associated particle fluxes, in particular those arising from heavy hadron decays.  In this work, we provide state-of-the-art QCD predictions for heavy hadron production including radiative corrections at next-to-leading order and  using parton distribution functions including small-$x$ resummation at next-to-leading logarithmic accuracy.  We match our predictions to parton showers to provide a  realistic description of hadronisation effects. We demonstrate the utility of our predictions by presenting the energy spectrum of neutrinos from charm hadron decays. Furthermore, we employ our predictions to estimate, for the first time, FASER's sensitivity to electrophilic ALPs, which are predominantly generated in beauty hadron decays.
}


\maketitle

\section{Introduction} 
\label{sec:introduction}

The main experimental program at the LHC traditionally focuses on particles with sizable transverse momentum in the central region, such as those expected to be produced in the decay of Standard Model (SM) bosons or proposed heavy new particles at the TeV scale.  More recently, there is also a growing interest in particles with small transverse momentum moving in the forward region. Specifically, forward hadrons are produced in enormous numbers and can inherit a substantial fraction of the beam energy. These hadrons can then decay into neutrinos, generating an intense and tightly collimated beam of high-energy neutrinos along the collision axis of the beams. Moreover, these forward hadrons might also decay into so-far undiscovered feebly interacting light particles, which have been posited by various models of new physics and may play the role of dark matter or be a mediator to the dark sector~\cite{Feng:2017uoz}.

Two new LHC experiments --- FASER~\cite{FASER:2022hcn} and SND@LHC~\cite{SNDLHC:2022ihg} --- have recently started their operation in the forward region to exploit this opportunity. Indeed, first direct observation of collider neutrinos was reported by FASER in March 2023~\cite{FASER:2023zcr} and shortly after also by SND@LHC~\cite{SNDLHC:2023pun}. In addition, FASER has performed a first search for dark photons yielding world-leading constraints~\cite{FASER:2023tle}. These experiments will operate during the third run of the LHC further studying collider neutrinos and searching for signs of new physics. Looking further into the future, the Forward Physics Facility (FPF) has been proposed to house a suite of experiments to continue this program during the high-luminosity LHC (HL-LHC) era~\cite{Anchordoqui:2021ghd, Feng:2022inv}.\sloppy

This emerging forward LHC search and neutrino program requires reliable estimates of the associated particle fluxes. In particular, this requires precise predictions of the forward hadron fluxes and their uncertainties. Since forward light hadrons are of non-perturbative origin, their production is conventionally simulated using dedicated event generators, most of which are developed for cosmic ray physics. In contrast, forward heavy charmed and beauty hadron production can in principle be described by perturbative QCD methods. While several such predictions exist in FASER kinematics, utilizing both Monte Carlo generators~\cite{Kling:2021gos} and analytic perturbative calculations~\cite{Bai:2020ukz, Bai:2021ira, Maciula:2022lzk, Bhattacharya:2023zei}, these often entail approximate descriptions of either the hard scattering or the hadronization that may affect their reliability. Indeed, it was noticed that their predictions for the forward neutrino flux differ by more than an order of magnitude. 

The use of state-of-the-art perturbative QCD predictions matched with parton showers for heavy quark production, which has been shown to provide a reliable description of high-rapidity LHCb data~\cite{Gauld:2015yia,Bertone:2018dse}, has so far never been  consistently employed in the very forward region probed in FPF kinematics. In this letter we build upon such framework in order to provide novel predictions for the expected forward neutrino event rate at FASER. Our results can be used to constrain a variety of New Physics models predicting feebly interacting particles produced in heavy meson decays. As an illustration, we will use our prediction to estimate for the first time FASER's sensitivity for electrophilic ALPs.

\section{Forward Hadron Production \\ at the LHC} 
\label{sec:production}

We start by introducing the theoretical framework used to obtain our predictions for forward heavy hadrons production.
We produce prediction at next-to-leading order (NLO) accuracy matched with Monte Carlo parton shower via the \textsc{Powheg} method~\cite{Nason:2004rx, Frixione:2007vw, Alioli:2010xd}.
The NLO calculation is performed in a fixed-flavour scheme with massive heavy quarks using the \texttt{hvq} generator~\cite{Frixione:2007nw}.
The fragmentation and the hadronisation of the heavy quarks are handled by the \textsc{Pythia} 8.2 shower~\cite{Sjostrand:2014zea}, including also the contribution from multi-parton interactions (MPI).

We use the NNPDF3.1sx+LHCb PDF set with $\alpha_s = 0.118$ at NLO+NLL$_x$ accuracy~\cite{Ball:2017otu,Bertone:2018dse} as our input set of parton densities through the LHAPDF interface~\cite{Buckley:2014ana}.
The motivation for this choice of PDF is twofold.
Firstly, we prefer to use a PDF set which includes LHCb $D$-meson production data~\cite{LHCb:2013xam,LHCb:2015swx,LHCb:2016ikn} to reduce the relevance of the PDF uncertainty at small values of the partonic $x$ probed in forward particle production~\cite{PROSA:2015yid,Gauld:2015yia,Gauld:2016kpd}.
Secondly, this PDF set includes small-$x$ (BFKL) resummation effects at NLL$_x$ (see \cite{Bonvini:2016wki,Bonvini:2017ogt} and references therein).
In light of the suggested evidence for an onset of BFKL dynamics at HERA data in the small-$x$ region~\cite{Ball:2017otu,xFitterDevelopersTeam:2018hym} we include small-$x$ resummation effects in our predictions in the forward region, which probe values of $x$ down to $x \lesssim 10^{-6}$.
We note that a consistent calculation should generally include small-$x$ resummation not only in the parton densities but also in the NLO partonic coefficient functions for heavy quark pair production, which however are not yet in a format amenable to LHC phenomenology.
Recent results for the production of a pair of beauty quarks at NLO+NLL$_x$~\cite{Silvetti:2022hyc} however show that at this order in $\overline{\rm MS}$-like schemes the bulk of the small-$x$ resummation effects is contained in the PDF evolution, whilst the impact on the partonic coefficient function is minor.
This hierarchy is valid also at relatively high values of the rapidity of the heavy quark pair, which probe smaller $x$ values. 
This suggests that the importance of resummation in the coefficient function is less relevant at NLO+NLL$_x$ even for charm production.
These results justify the approximation used in this work, where we include resummation effects only in the parton densities whereas we neglect the resummation effects in the partonic coefficient functions for heavy quark production.
For transparency, however, we shall denote our predictions as NLO+NLL$_x^{\rm PDF}$, to indicate that the resummation is included only in the PDFs.

\begin{figure*}[t]
    \includegraphics[width=0.49\textwidth]{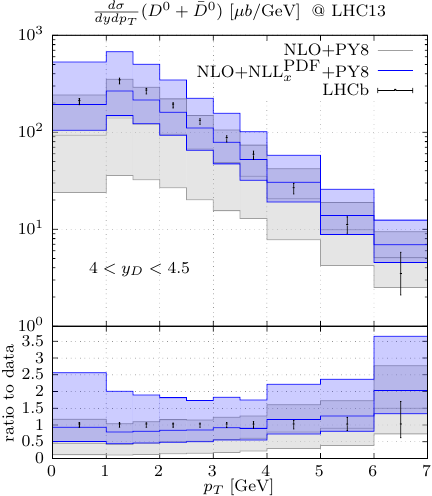}
    \includegraphics[width=0.49\textwidth]{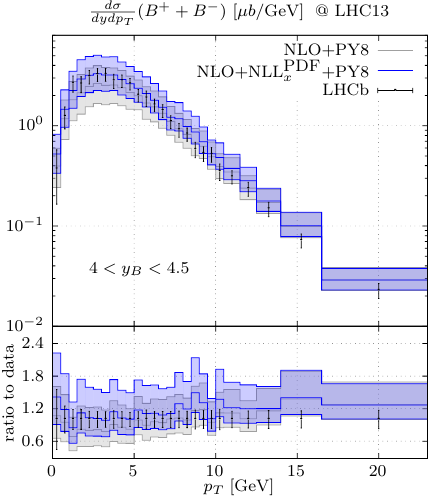}
    \caption{Predictions for the production of $D$-meson (left) and $B$-meson (right) in comparison with LHCb.}
    \label{fig:production}
\end{figure*}

We set the central factorisation and resummation scales equal to $\mu_F = \mu_R = (m_{Q}^2+p_{T,{Q}}^2)^{1/2}$ and the nominal beauty and charm quark mass to $4.5$ GeV and to $1.5$ GeV, respectively\footnote{We note that the beauty mass is set to 4.92 GeV in the NNPDF PDF set. Our choice for $m_b=4.5$~GeV is motivated by an improved description of the $B$ meson production data for our central prediction. Predictions obtained with a larger value of $m_b=4.92$~GeV are largely included in our scale uncertainty but provide a somewhat worse description of the LHCb data.}.
We shower the events with \textsc{Pythia} 8.245, using the default Monash tune~\cite{Skands:2014pea}.
We note that our approach only includes small-$x$ resummation effects inclusively (i.e. before showering).
A completely consistent treatment of small-$x$ resummation in predictions matched with parton shower is currently beyond the state of the art \footnote{Event generators which implement backward small-$x$ evolution such as CASCADE~\cite{Jung:2000hk,CASCADE:2010clj} are available.}.
As a validation of the robustness of our predictions, we show a comparison against LHCb data at 13 TeV for $D_0+\bar D_0$~\cite{LHCb:2015swx} and for $B^+ + B^-$~\cite{LHCb:2017vec} in the $4 < y_{D (B)} < 4.5$ window.
We observe that our NLO+NLL$_x^{\rm PDF}$ predictions provide an excellent description of the LHCb data, both at low and at high $p_T$.
For comparison, we show also predictions at NLO accuracy, obtained with a baseline NLO PDF set which is extracted using the same data as NNPDF3.1sx+LHCb, but does not include BFKL resummation effects.
The latter predictions still provide a good description of the data within uncertainties but tend to undershoot them, especially at low $p_T$.
The scale uncertainties are rather large, at the $30\%-40\%$ level for beauty and even larger for charm production.
Although we limit ourselves to the most forward bins, we note a similarly good description of the data at central rapidities, where the differences between NLO and NLO+NLL$_x^{\rm PDF}$ predictions become smaller.

We have verified that other sources of uncertainties such as PDF uncertainty, sensitivity to the quark mass used in the calculation, as well as variation of the \textsc{Pythia} tune are below the scale uncertainty.
We stress that, especially at large rapidities, the inclusion of LHCb forward data in the PDFs is instrumental in reducing the PDF uncertainty.
We note the use of alternative colour reconnection schemes, e.g.~\cite{Christiansen:2015yqa}, improves the baryon enhancement towards low $p_T$ observed experimentally~\cite{ALICE:2017thy, ALICE:2020wla, ALICE:2021rzj}, but does not affect our predictions shown below.
We also checked that the recently presented forward tune~\cite{Fieg:2023kld}, which improves the modelling of beam remnant hadronization, has a minor impact compared to the scale uncertainties.

Finally, we have also considered a different shower, by matching the \textsc{Powheg} predictions to \textsc{Herwig} 7.2~\cite{Bewick:2019rbu,Bellm:2019zci}, using the interface developed in Ref.~\cite{FerrarioRavasio:2018whr}.
We note that the two showers use different hadronisation models; PYTHIA uses string fragmentation~\cite{Andersson:1983ia} while HERWIG uses cluster fragmentation~\cite{Marchesini:1983bm}.
We find that, using the same setup as for the  \textsc{Powheg+Pythia} case, \textsc{Powheg+Herwig} provides a similar description of the LHCb data after hadronisation and multi-parton interactions, with a deterioration of the agreement with the data at higher values of the transverse momentum. Since it appears that \textsc{Powheg+Herwig} offers a somewhat worse description of the LHCb data, albeit compatible within the large scale uncertainties with the \textsc{Powheg+Pythia} predictions, we use \textsc{Powheg+Pythia} as our default prediction without considering the \textsc{Powheg+Herwig} results as an additional source of  uncertainty. The comparison between \textsc{Powheg+Herwig} and LHCb data is shown in Appendix~\ref{AppendixShower}.
The treatment of forward charm production requires special care due to the challenges in the description of forward charm hadronization (see e.g. Sect. 6.2.2 of Ref.~\cite{Feng:2022inv}).
The compatibility between our results, obtained with different tunes within~\textsc{Pythia} and with a different hadronisation model through \textsc{Herwig}, supports the robustness of the modelling of charm hadronization. A more comprehensive study of forward charm hadronization would further corroborate our results, but such a study goes beyond the scope of this work.

The large scale uncertainties could be reduced by increasing the perturbative accuracy of the calculation.
NNLO+PS accurate predictions for $B$-mesons have been shown to agree well with experimental data, with scale uncertainties reduced by more than a factor of two~\cite{Mazzitelli:2023znt}.
Nevertheless, at this accuracy other sources of uncertainty, most notably the PDF uncertainty and uncertainties related to the shower settings, should be assessed as they start to become comparable to the missing higher order uncertainties, especially at large rapidities.
Moreover, due to the values of the partonic $x$ probed, it would be necessary to investigate the impact of small-$x$ resummation on top of the NNLO correction; at this order, it should be assessed whether it is legit to neglect the effect of resummation in the partonic coefficient function, which would require the ingredients for a NNLO+NLL$_x$ matching, which are currently not fully available.
Finally, let us mention that a fully consistent treatment would require, alongside the resummation of small-$x$ logarithms, the joint resummation of small and large-$x$ logarithms, both in the coefficient function and in PDFs.
Currently such predictions are available only for very inclusive processes~\cite{Bonvini:2018ixe}, while a PDF set which include both small- and large-$x$ effects is not available.
The large scale uncertainties at NLO+NLL$_x^{\rm PDF}$ are sufficiently conservative to neglect the interplay of large and small-$x$ resummation.

For these reasons, in this work we prefer to use NLO+NLL$_x^{\rm PDF}$ accurate predictions, as the scale uncertainties quoted at this order provide a conservative estimate of the theoretical error on heavy quark production in the forward region.

\medskip 

In contrast to forward heavy hadron production, light hadron production cannot be described reliably by perturbative QCD due to the small values of $Q \lesssim 1$~GeV probed.
Moreover, a description based on a fragmentation function approach may not be appropriate due to the interplay with the beam remnants, see e.g.\cite{Fieg:2023kld}.
Instead, light hadron production is typically described by hadronic interaction models. 
In this work, we use several Monte Carlo event generators developed for cosmic ray physics but also commonly used to describe forward particle  production: \textsc{Epos-Lhc}~\cite{Pierog:2013ria}, \textsc{Sibyll~2.3d}~\cite{Riehn:2019jet} and \textsc{QgsJet~2.04}~\cite{Ostapchenko:2010vb} as implemented in the CRMC interface~\cite{CRMC}. 
The predictions for these generators have been validated against LHCf data for forward photons and neutrons at the 13~TeV collision energy~\cite{LHCf:2017fnw, LHCf:2020hjf} and form an envelope around the data.
When presenting results, we will use \textsc{Epos-Lhc} to obtain our central prediction and use the spread of the three generator predictions as an estimate of the flux uncertainty\footnote{We note that an alternative definition of uncertainties, using tuning variations in Pythia, has been proposed in \cite{Fieg:2023kld}. The study found that the uncertainties obtained this way are similar to those obtained using the spread of generators.}.

\section{Application at FASER and FPF} 
\label{sec:FASER}

Having discussed our predictions for forward heavy hadron production, let us now turn to their application in current and future forward physics experiments. These experiments utilize that the forward hadrons may decay into neutrinos, or potentially even into as-yet-undiscovered light dark sector particles, and hence create an intense, strongly-focused, and highly energetic beam of these particles along the beam collision axis.

One of these experiments is FASER, which is situated about 480m downstream of the ATLAS interaction point in a previously unused side tunnel of the LHC~\cite{FASER:2022hcn}. FASER is aligned with the beam collision axis and covers pseudorapidities $\eta \gtrsim 9$. Located at its front is the FASER$\nu$ neutrino detector, which  consists of a $25~\cm \times 25~\cm \times 1~\m$ tungsten target with roughly 1.2 tons target mass that is interleaved with emulsion films~\cite{FASER:2019dxq, FASER:2020gpr}. This detector provides a high resolution image of the charged particle tracks produced in neutrino interaction and allows the identification of the neutrino flavor as well as the measurement of their energy~\cite{FASER:2021mtu, FASER:2868284}.  Located behind is FASER's long-lived particle detector~\cite{FASER:2018ceo, FASER:2018bac, FASER:2018eoc}. It consists of a cylindrical decay volume with 1.5~m length and 10~cm radius, which is preceded by a veto system and followed by a spectrometer and a calorimeter. It is optimized for particle decays into electron pairs, for which it was found to have a good acceptance and negligible background~\cite{FASER:2023tle}. 

Upgraded detectors to continue the forward physics program are envisioned for the HL-LHC era. These would be housed within the proposed FPF, a dedicated cavern to be constructed 620~m downstream of ATLAS and designed to accommodate a suite of experiments~\cite{Anchordoqui:2021ghd, Feng:2022inv}. This proposal encompasses three neutrino detectors as well as FASER2 for long-lived particle searches and FORMOSA for milli-charged particles searches~\cite{Foroughi-Abari:2020qar}. While different designs have been considered for FASER2, we assume it to consist of a 1~m radius and 10~m long cylindrical decay volume. 

In the following, we employ our results on forward heavy hadron production to predict neutrino fluxes arising from charm decay and the search sensitivity for electrophilic ALPs at FASER. We emphasize that these predictions are also applicable to other experiments at the FPF and other physics contexts beyond the considered models. 

\subsection{Neutrinos}
\label{sec:neutrinos}

\begin{figure*}
   \centering
  \includegraphics[width=0.49\linewidth]{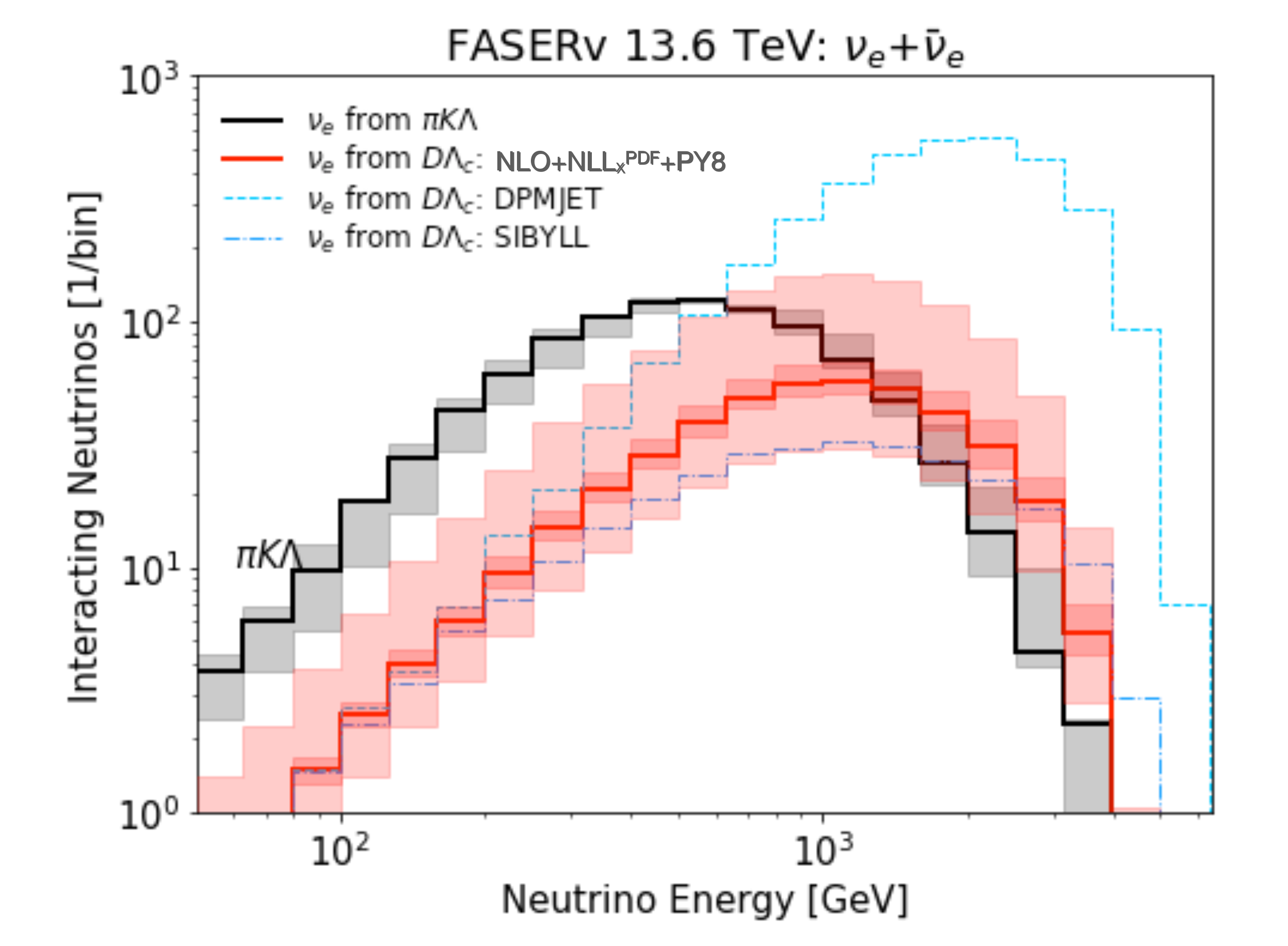}
  \includegraphics[width=0.49\linewidth]{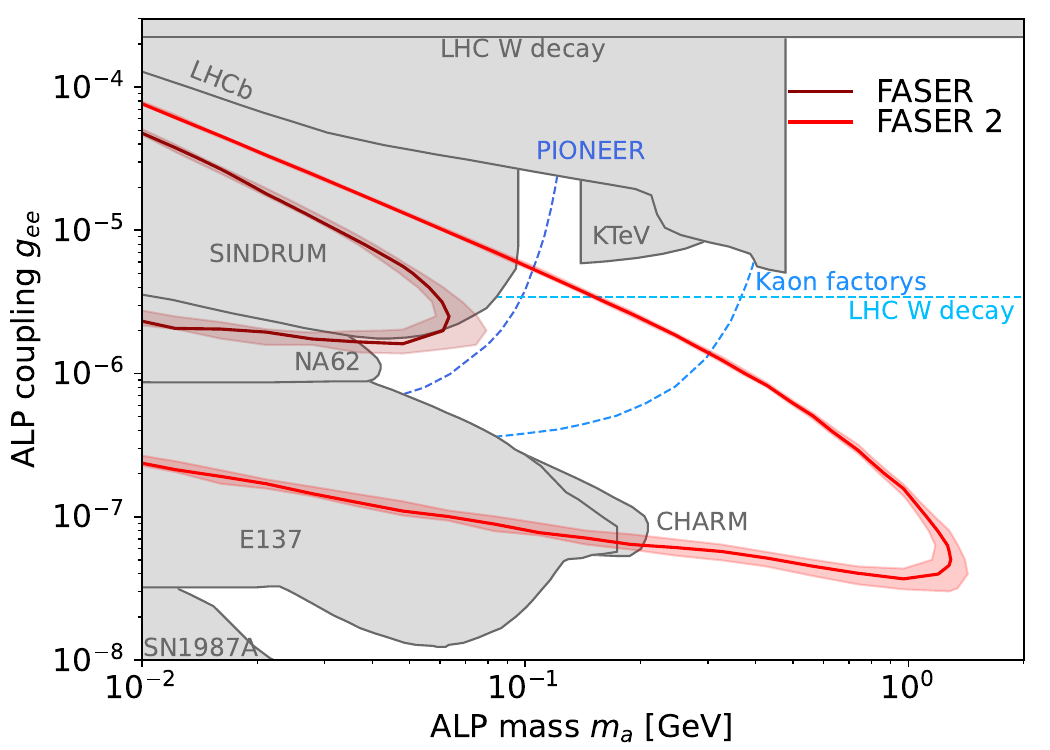}
  \caption{\textbf{Left:} Predicted energy spectrum of electron neutrinos from charm hadrons decay at FASER$\nu$. We show the central prediction as red solid line, the associated uncertainty as light shaded band. For reference, we also display PDF uncertainties in a darker shade of red. Alternative predictions obtained with \textsc{Sibyll~2.3d} and \textsc{DpmJet~3.2019.1} are shown as blue dashed lines. The neutrino component from light hadron decays is shown in grey. 
  \textbf{Right:} Sensitivity of FASER during LHC Run3 with $200~\ifb$ and FASER2 at the HL-LHC with $3~\iab$ in the ALP parameter space. The solid lines correspond to the central prediction of the production rate, while the shaded bands represent the production uncertainty. Existing constraints are shown as grey shaded regions and the blue dotted lines show the expected sensitivity of future experiments. 
  }
  \label{fig:results}
\end{figure*} 

One of FASER's main objectives is the study of high energy collider neutrinos. The forward-moving neutrinos traversing FASER primarily originate from the weak decay of the lightest mesons and baryons associated with a specific flavor, notably pions, kaons, and charm hadrons. As discussed in Ref.~\cite{Kling:2021gos}, charm hadron decays are expected to contribute predominantly to electron and tau neutrinos, while muon neutrinos and low-energy electron neutrinos mainly stem from light hadron decays.
The component arising from $B$ meson is subdominant, and is discussed in Appendix~\ref{sec:app_fromB}.

We employ our derived results for forward charm hadron production to predict the expected number of neutrino interactions within FASER$\nu$. For this, we consider the configuration of the LHC used at the beginning of Run~3 with a 13.6~TeV center-of-mass energy and a beam half-crossing angle of 160~$\mu$rad downwards. To compute the expected events in FASER$\nu$, we fold the neutrino flux with interaction cross-sections obtained from \textsc{Genie}~\cite{Andreopoulos:2009rq}.  We note that the Bodek-Yang model employed in \textsc{Genie} agrees with more recent predictions, and that cross-sections uncertainties are generally much smaller than the flux uncertainties~\cite{Candido:2023utz}.

The outcome is depicted in the left panel of Fig.~\ref{fig:results}, which presents the energy spectrum of interacting electron neutrinos originating from charm hadron decay within FASER$\nu$ with $200~\ifb$. The solid red line represents the central prediction, while the shaded band corresponds to the associated scale uncertainties. These uncertainties approximately result in a factor of two variation, which is roughly constant across the energy range. For reference, we also display the much smaller PDF uncertainties in a darker shade of red. For comparative purposes, we also display predictions based on \textsc{Sibyll~2.3d} and \textsc{DpmJet~3.2019.1}~\cite{Roesler:2000he, Fedynitch:2015kcn}. Our prediction is relatively close to the former, while the latter yields notably larger flux predictions for neutrinos originating from charm decay, respectively.

In addition, to guide the eye, we also show the event rate arising from light hadron decays as a grey band, which was obtained using the fast neutrino flux simulation introduced in Ref.~\cite{Kling:2021gos}. We can see that the neutrino flux component from charm decay will provide the leading contribution for electron neutrinos with energies above roughly 1~TeV. Overall, we predict  $439^{+757}_{-201} \ \nu_e$, $426^{+733}_{-194} \ \nu_\mu$ and $25^{+40}_{-11} \ \nu_\tau$  charged current neutrino interactions from charm hadron decays to occur in FASER$\nu$ during Run~3 with $200~\ifb$.
%

\subsection{Electrophilic ALPs} 
\label{sec:ALPs}

FASER's other primary objective is the search for light long-lived particles as predicted by many models of new physics. One prominent example are axion-like particles (ALPs). Multiple ALP-models have been studied in the context of FASER, such as ALPs with dominant coupling to photons~\cite{Feng:2018pew}, to gluons and quarks~\cite{FASER:2018eoc} and to weak gauge bosons~\cite{Kling:2020mch}. Here we consider another, yet unexplored, possibility: an ALP with dominant couplings to electrons, colloquially referred to as electrophilic ALP. 
 
Following Ref.~\cite{Altmannshofer:2022izm}, we consider a scenario in which the interaction of electrophilic ALPs with the SM is described by the interaction Lagrangian $\mathcal{L} = \frac{g_{ee}}{2 m_{e}} \, \partial_{\mu} a \ \bar{e} \gamma^\mu \gamma_5 e$.  In addition to its couplings to electrons, such an ALP also acquires couplings to the weak gauge bosons and photons through the chiral anomaly~\cite{Altmannshofer:2022izm}. The most relevant implication is that, through the $W$-boson coupling, the electrophilic ALP can be produced in flavor-changing hadron decays. Indeed, in the forward region of the LHC, the dominant production channel of such electrophilic ALPs are rare $B$-meson decays $B \to X_s a$ as well as kaon decays $K \to \pi a$. In addition, we also consider three-body meson decays of the type $P^\pm \to e \nu a$ for $P = \pi, K, D$ and $ D_s$. A detailed overview over the electrophilic ALP model and its phenomenology can be found in App.~\ref{AppendixALPs}.

In the considered MeV to GeV mass range, the electrophilic ALP mainly decays into electron pairs, with decays into photon pairs also becoming important at higher masses. Notably, for sufficiently small couplings $g_{ee}$, the ALP becomes long-lived, allowing it to travel a macroscopic distance before decaying, for example, in FASER. In the following we assume that FASER and FASER2 can detect the signal with full efficiency and negligible backgrounds~\footnote{This assumption is certainly justified for the considered di-electron signature for FASER, see Sect.~\ref{sec:FASER}. For the projected FPF reach such an assumption will generally depends on the capabilities of the proposed detector to distinguish the di-electron signal from possible neutrino signatures. The detector is currently being designed, with background rejection being one of the main criteria.}.  

To determine FASER's sensitivity to electrophilic ALPs, we incorporate the model characteristics into the \textsc{Foresee} package~\cite{Kling:2021fwx}. The resulting reach, which corresponds to three signal events in the detector, for FASER during LHC Run3 with $200~\ifb$ and FASER2 at the HL-LHC with $3~\iab$ in the ALP parameter space spanned by its coupling $g_{ee}$ and mass $m_{a}$, is shown in the right panel of Fig.~\ref{fig:results}. The solid lines represent the central prediction, while the shaded bands reflect the production uncertainty introduced in Sec.~\ref{sec:production}. We note that, despite the substantial flux uncertainties, their overall impact on the sensitivity reach remains relatively small due to a strong coupling dependence at both small and large couplings. The flux uncertainties predominantly affect the reach at the high-mass end of the sensitivity region. 

The grey regions have previously been constrained using searches for long-lived particles at E137~\cite{Bjorken:1988as} and CHARM~\cite{CHARM:1985anb}; rare $B$-meson decays at LHCb~\cite{LHCb:2015ycz}; rare kaon decays at NA62~\cite{NA62:2020xlg} and KTeV~\cite{KTeV:2003sls};  rare pion decays at SINDRUM~\cite{SINDRUM:1989qan}; and rare W boson decays~\cite{Altmannshofer:2022izm} as well as supernova SN-1987A~\cite{Lucente:2021hbp}. The blue dashed lines indicate the potential future sensitivity of searches for rare pion decays at PIONEER~\cite{PIONEER:2022yag}, rare kaon decays at kaon factories~\cite{Goudzovski:2022vbt}, and rare W decays at the LHC~\cite{Altmannshofer:2022izm}. The sensitivity of searches at future colliders has also been studied~\cite{Lu:2022zbe}. All shown bounds and potential sensitivities were taken from Ref.~\cite{Altmannshofer:2022izm}. FASER will independently constrain part of the ALP parameter space only been assessed by a reinterpretation of the SINDRUM measurement, but barely probe unexplored parameter space at the end of LHC Run 3. In contrast, FASER2 will extend this reach drastically, and be able to probe yet unconstrained parameter-regions up to ALP masses of 1~GeV. Noticeably, it will probe regions not projected to be probed by any other experiment.

\section{Conclusions}
\label{sec:conclusions}

Measurements of neutrinos and searches for feebly interacting particles at the LHC are attracting growing interest thanks to the construction of two new experiments probing the very forward region. This physics program may become even more relevant with the envisioned future Forward Physics Facility which could start operating during the high luminosity phase of the LHC. 

In this context, it is of central importance to provide reliable estimates for the relevant particle fluxes and their associated uncertainties, which, in particular, entail heavy (light) hadron production. In this letter, we present new predictions for forward heavy hadron production in the FASER kinematics, based on state-of-the-art QCD calculations. Our predictions combine the NLO radiative corrections with the effective inclusion of small-$x$ resummation at NLL, and are matched to the \textsc{Pythia} parton shower program to provide a realistic description of hadronisation effects. 

We use our results for two relevant applications at FASER: i) the reliable prediction of neutrino fluxes in the forward region, and ii) the sensitivity to long lived particles arising in new physics scenarios. We find that, despite the relatively large uncertainties, our predictions for the energy spectrum of interacting neutrinos coming from charmed hadrons disfavour some of the results obtained with other less accurate frameworks. In the case of long lived particles, we focus on an electrophilic ALP scenario. We find that the sensitivity reach of FASER is competitive and complementary to existing bounds, while the FASER2 upgrade will explore a substantially larger region of the parameter space. 

The predictions for forward hadron production from this study will open the door to numerous additional applications, including the use of LHC neutrino flux measurements to probe QCD in novel kinematic regimes~\cite{Kling:2023tgr} and of high-energy neutrino scattering to investigate into nuclear structure~\cite{Cruz-Martinez:2023sdv}. Furthermore, they will enhance the sensitivity of forward experiments in the pursuit of new physics.
The large perturbative uncertainties characterising forward heavy hadron production are expected to decrease significantly thanks to the present (near future) availability of NNLO predictions for bottom (charm) production matched with parton showers. It will thus become crucial to reduce the other source of uncertainty, currently subdominant, notably hadronisation and PDF uncertainty. In this respect, it may be beneficial to exploit data collected at FASER and SND@LHC to reduce these uncertainties, by tackling the challenging aspects related to the reconstruction of the underlying QCD heavy-quark production from the neutrino scattering events measured in forward detectors.

  Our results are publicly available at the \texttt{GitHub} repository \url{https://github.com/lucarottoli/forward_heavy_hadrons_NLONLLx}.


\section*{Acknowledgements}

We thank Jeff Dror, Rhorry Gauld, Markus Prim, Geraldine Servant and the members of the FPF forward charm production working group for many fruitful discussions. We are grateful to the authors and maintainers of many open-source software packages, including \textsc{Rivet}~\cite{Buckley:2010ar, Bierlich:2019rhm} and \textsc{scikit-hep}~\cite{Rodrigues:2020syo}. L.B. is supported by the Swiss National Science Foundation (SNSF) under contract 200020 188464. F.K. acknowledges support by the Deutsche Forschungsgemeinschaft under Germany's Excellence Strategy - EXC 2121 Quantum Universe - 390833306.  L.R. is supported by the SNSF under contract PZ00P2 201878. 

\appendix

\section{Choice of parton shower settings}
\label{AppendixShower}

\begin{figure*}[t]
    \includegraphics[width=0.49\textwidth]{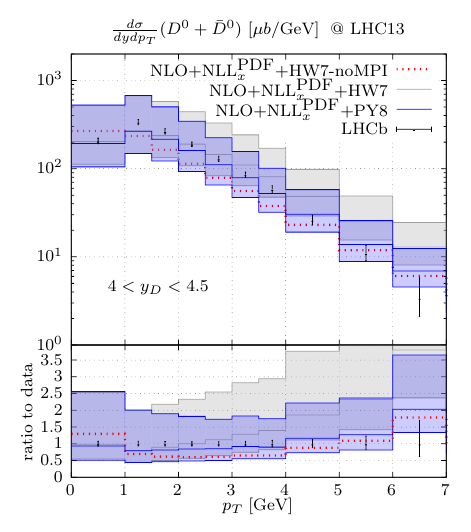}
    \includegraphics[width=0.49\textwidth]{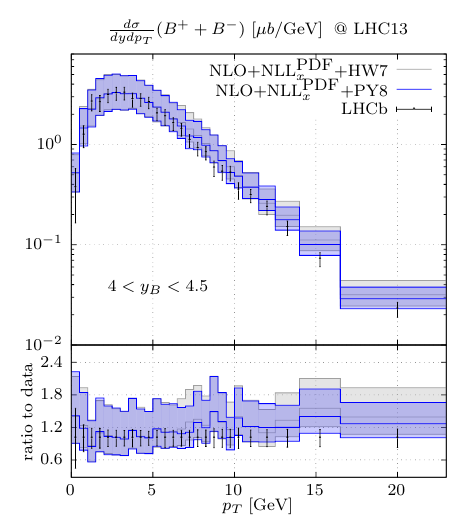}
    \caption{Predictions for the production of $D$-meson (left) and $B$-meson (right) with \textsc{Powheg+Herwig}, compared with our default results using \textsc{Powheg+Pythia} and the LHCb data. In the case of $D$-meson production we show also the results obtained with  \textsc{Powheg+Herwig} without the inclusion of MPI.}
    \label{fig:productionHerwig}
\end{figure*}

\begin{figure*}[t]
    \includegraphics[width=0.49\textwidth]{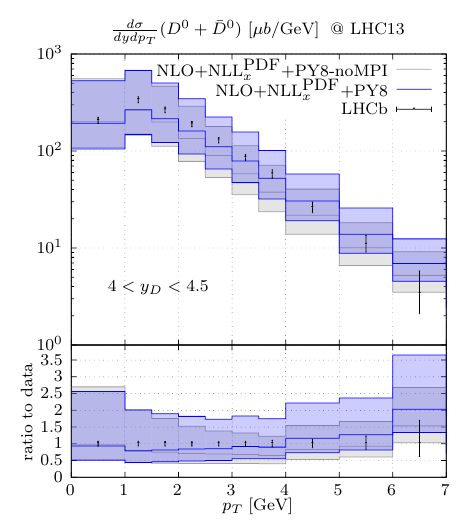}
    \includegraphics[width=0.49\textwidth]{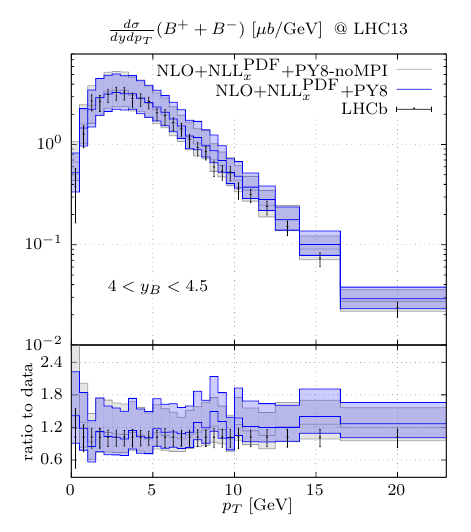}
    \caption{Predictions for the production of $D$-meson (left) and $B$-meson (right) with \textsc{Powheg+Pythia} with and without the inclusion of MPI effects compared to the LHCb data.}
    \label{fig:productionMPI}
\end{figure*}

In this appendix we compare the default predictions for forward beauty and charm production obtained with \textsc{Powheg+Pythia} with the ones obtained matching our \textsc{Powheg} results to the \textsc{Herwig} parton shower program.
We also comment on the effect of the removal of multi-parton interactions from our default \textsc{Powheg+Pythia} setup.
We also briefly comment on the use of a different shower model within \textsc{Pythia}.

\begin{figure}[t]
    \includegraphics[width=0.49\textwidth]{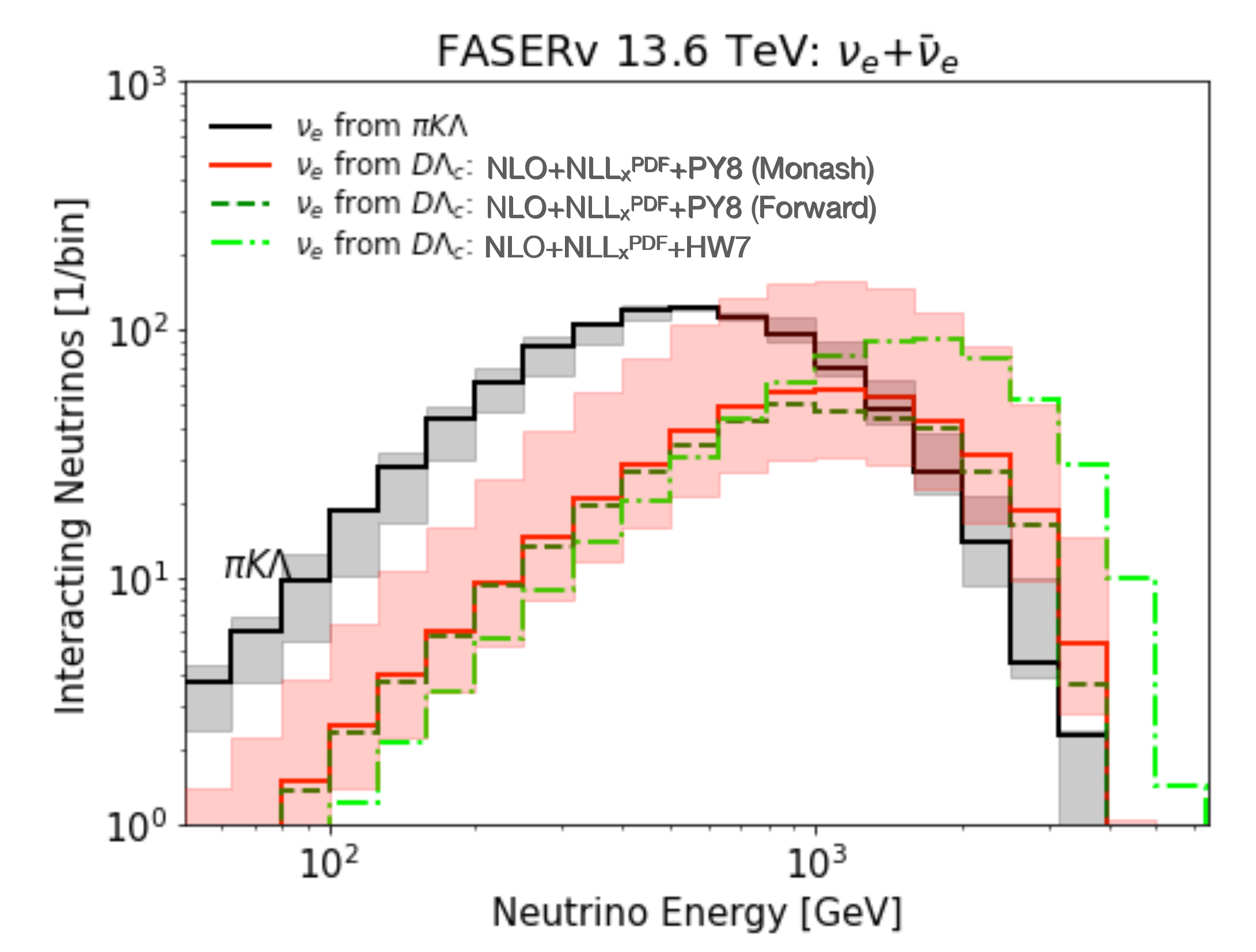}
    \caption{Same as Fig.~\ref{fig:results} left, now comparing different \textsc{Pythia} tunes as well as the results obtained using \textsc{Powheg+Herwig}.}
    \label{fig:fluxesvariations}
\end{figure}

We start by comparing the predictions of \textsc{Powheg+Herwig} to those of \textsc{Powheg+Pythia}.
The comparison is shown in Fig.~\ref{fig:productionHerwig} for charm (left panel) and beauty (right panel) production in the rapidity window $4 < y < 4.5$.
We observe an overall agreement between the two results within the large scale uncertainties, especially at low values of the transverse momentum of the mesons.
At higher transverse momentum the spectrum obtained with \textsc{Powheg+Herwig} is harder; the effect is relatively mild in the case of beauty meson production, while it is larger in the case of charm meson production.
In particular, the last two bins of the \textsc{Powheg+Herwig} distribution overshoots the data, which end outside the relatively large scale uncertainty bands of the NLO prediction.
This effect is present also at lower meson rapidities, leading to an overall worse description of the LHCb charm data when using \textsc{Powheg+Herwig}.

We noticed that such trend is alleviated when removing the effect of multi-parton interactions from our \textsc{Powheg+Herwig} predictions (see dashed red curve in the left panel of Fig.~\ref{fig:productionHerwig}).  
However, the removal of MPI leads to distortion of the shape of the distribution towards low values of the meson transverse momentum.
Without MPI, \textsc{Powheg+Herwig} does not seem to reproduce the presence of a peak in the experimental distribution of $D$-mesons for $p_T \simeq 1-2$ GeV and tends to undershoot the data between 2 and 5 GeV, especially in the central rapidity bins.

In contrast, the effect of MPI is milder in the case of \textsc{Powheg+Pythia} predictions, as we show in Fig.~\ref{fig:productionMPI}.
For beauty meson prediction the removal of MPI effects leads only to minor differences in the two spectra.
For charm meson prediction the inclusion of MPI leads to a somewhat harder spectrum, which offers a partially improved description of the data, especially after the peak of the distribution.
Also in this case we observe that the inclusion of MPI effects leads to a harder spectrum, but the description of the data in the tail is improved with respect to the \textsc{Powheg+Herwig} case.

Finally, in Fig.~\ref{fig:fluxesvariations} we show our results for the energy spectrum of interacting electron neutrinos originating from charm hadron, as in Fig.~\ref{fig:results}, using the default MONASH tune, the forward tune~\cite{Fieg:2023kld} and \textsc{Powheg+Herwig}.
We observe that the effect of using another tune is minor, whereas the use of a different parton shower has a somewhat larger effect especially at small and large values of the neutrino energy, although the uncertainty bands (not shown for the \textsc{Powheg+Herwig} result) do overlap.
Since the predictions obtained with \textsc{Powheg+Herwig} provide a somewhat worse description of the forward data for $B$ and $D$ meson production, we prefer not to increase our uncertainty bands, as the differences between the two parton showers may well be reduced with a proper tuning of \textsc{Herwig}.

We have also checked whether the use of a different shower model within \textsc{Pythia} significantly modifies the \textsc{Powheg+Pythia} results. Specifically, we have considered the use of VINCIA~\cite{Giele:2013ema} and DIRE~\cite{Hoche:2015sya} shower models. We found that the differences with respect to the default \textsc{Pythia} shower can be as large as those between \textsc{Powheg+Pythia} and \textsc{Powheg+Herwig}, although within the quoted uncertainties. Moreover, the differences are affected by the interplay with the shower model and the specific tune settings used in VINCIA and DIRE. For this reason, we refrain from showing the results obtained with these shower models as more work is needed to study such an interplay.

\section{Electrophilic ALPs}
\label{AppendixALPs}

\begin{figure*}
  \centering
  \includegraphics[width=0.49\textwidth]{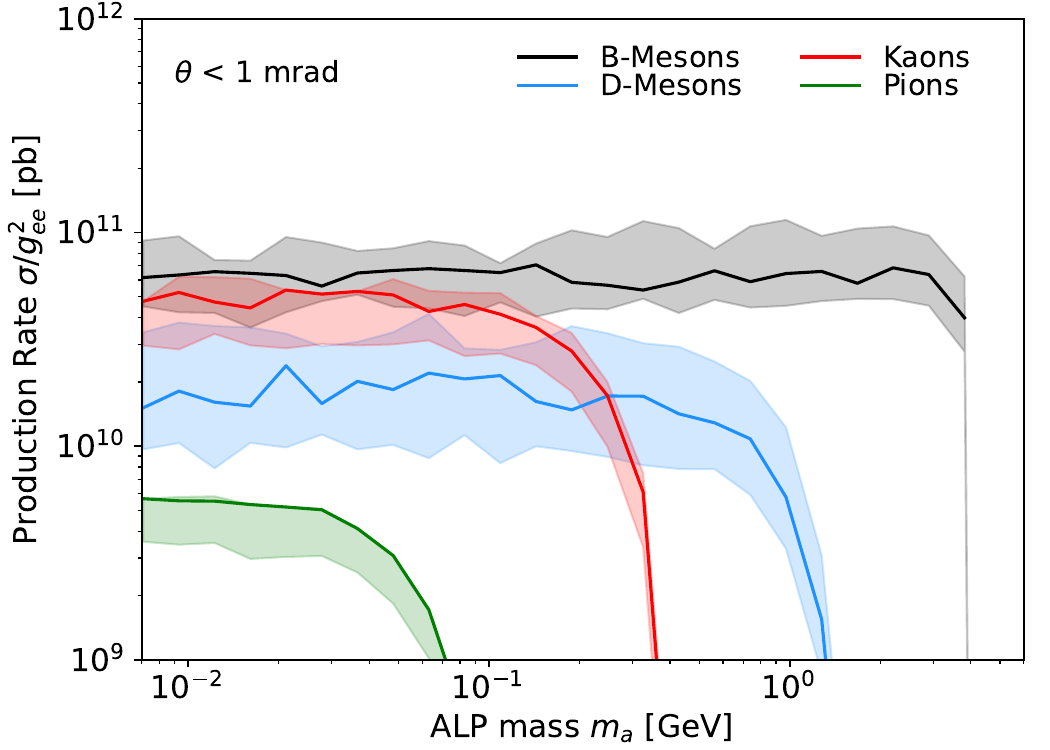}
  \hfill
  \includegraphics[width=0.49\textwidth]{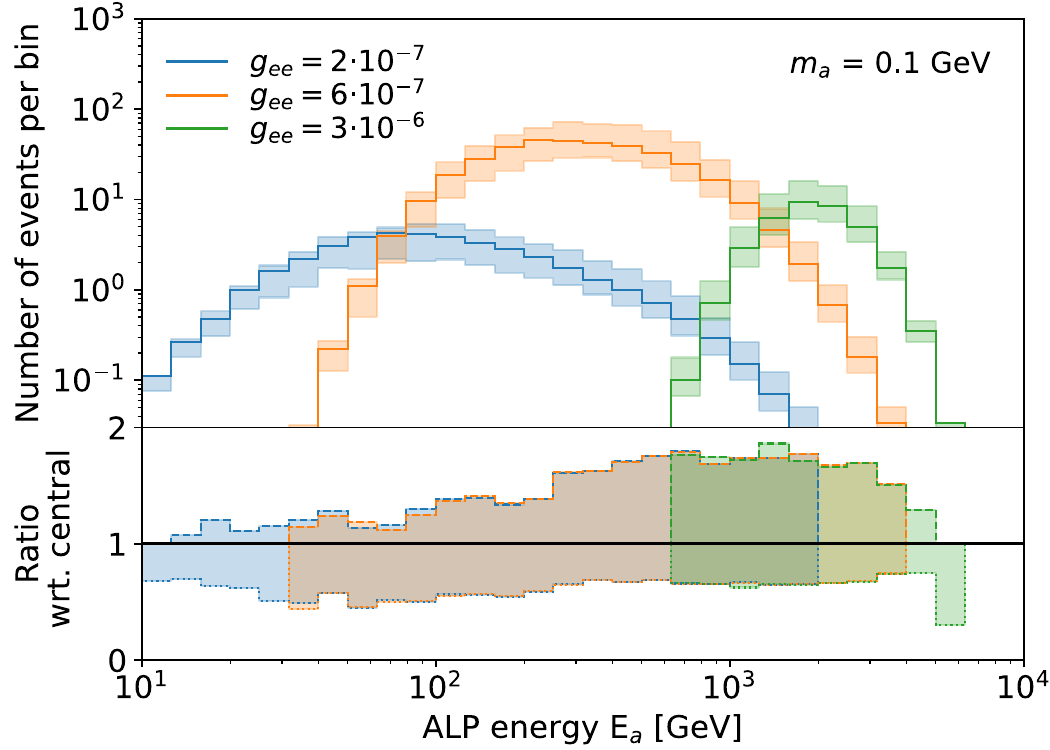}
  \caption{\textbf{Left:} ALP production rate via decay of various mesons as a function of ALP mass $m_a$ at a angular acceptance $\theta < 1$~mrad. The lines show the sum of all production channels of the respective meson. The uncertainty band was derived by varying the scales for charm and beauty mesons, and by varying the generators for pions and kaons as discussed in Sec.~\ref{sec:production}. \textbf{Right:} Expected energy spectrum of ALPs decaying in the FASER2  decay volume for three different ALP benchmark models. The shaded band corresponds to the flux uncertainty. The lower panel shows the same flux normalized by the central predictions.} 
  \label{fig:ProdRates}
\end{figure*}

In this appendix, we provide more details on the electrophilic ALP model and its phenomenology. As mentioned in the main text, the Lagrangian describing the electrophilic ALPs interaction with the SM is given as~\cite{Altmannshofer:2022izm}
\be
\mathcal{L}_{a}= \partial_{\mu} a   \frac{g_{e e}}{2 m_{e}} \bar{e} \gamma^\mu \gamma_5 e.
\ee
After integration by parts and considering the chiral anomaly, this can be written as
\be
\!\!\mathcal{L}_{a} \!=\!  ag_{e e}\bigg(\bar{e} i \gamma_{5} e+\frac{e^{2}}{16 \pi^{2} m_{e}}\left[\frac{1}{4 s_{W}^{2}} W_{\mu \nu}^{+} \tilde{W}^{-, \mu \nu}\right.  \\
-F_{\mu \nu} \tilde{F}^{\mu \nu} \bigg] +\frac{i g}{2 \sqrt{2} m_{e}}\bar{e} \gamma^{\mu} P_{L} \nu W_{\mu}^{-}+\text {...}\bigg)\ . 
    \label{Eq:Lagrangian}
\ee
Notably, in addition to its coupling to electrons, the ALP also obtains couplings to the photon and $W$-boson. Additional couplings to $ZZ$ and $Z\gamma$ also exist, but are not shown since they are not relevant for this work. \medskip

\noindent \textbf{ALP Production at the LHC:} At the LHC, the electrophilic ALP can be produced in both two-body hadron decays and three-body meson decays. For the two-body decays, we find the primary production channels to be kaon and $B$-meson decays. Using the results obtained in Ref.~\cite{Altmannshofer:2022izm}, the corresponding branching fractions are:
\be
\text{BR}_{K^{\pm} \rightarrow \pi^{\pm} a} &= 45 \times g^2_{ee} \times \lambda^{1 / 2}_{m_{K^+},m_{\pi^+},m_a}\, ,\\
\text{BR}_{K_L \rightarrow \pi^0 a} &= 27 \times g^2_{ee} \times \lambda^{1 / 2}_{m_{K^0},m_{\pi^0},m_a}\, , \\
\text{BR}_{K_S \rightarrow \pi^0 a} &= 0.3 \times g^2_{ee} \times \lambda^{1 / 2}_{m_{K^0},m_{\pi^0},m_a}\, , \\
\text{BR}_{B \rightarrow X_sa} &= 1.6 \cdot 10^5\times g_{ee}^2 \times \lambda_{m_B,0,m_a} \, . 
\ee
with the K\"all\'en function
\be 
\lambda_{abc}\!=\!\frac{{a}^4\!+\!{b}^{4}\!+\!{c}^{4}\!-\!2({a}^2{b}^2\!+\!{a}^2{c}^2\!+\!{b}^2{c}^2)}{{a}^4}.
\ee
In particular, for production via $B$-meson decay, here we follow the spectator model approach presented in Ref. \cite{Winkler:2018qyg}. A comparison with other approaches is shown in App.~\ref{sec:app_b}. 

ALPs can also be produced in three-body decays of the type $P \to e \nu_e a$, where $P$ is a pseudoscalar meson. For this, we use the differential decay width 
\be
\frac{d\text{BR}(P^\pm \to e \nu_e a)}{dE_a}=C_P \, g_{ee}^2  (E_a^2-m_a^2)^{\frac{3}{2}},
\ee
where 
\be
C_P = \frac{\text{BR}(P^\pm \rightarrow \ell^{\prime+} \nu_{\ell^{\prime}})}{24\pi^{2}m_{e}^{2}m_{\ell^{\prime}}^{2}} \lambda^{-1}_{P,\ell^{\prime},0},
\ee
which was obtained from the results presented in Ref.~\cite{Altmannshofer:2022izm}. For the coefficients, we obtain $C_{\pi} = 7.6 \times 10^{6}~\gev^{-4}$, $C_{K} = 9.9\times 10^{4}~\gev^{-4}$, $C_{D_s} = 7.9\times 10^{3}~\gev^{-4}$, and $C_{D} = 5.5\times 10^{2}~\gev^{-4}$. We find that the most relevant three-body decay channels are those of kaons and $D_{s}$-mesons. 

The production rate of ALPs within 1~mrad around the beam collision axis as a function of ALP mass  is shown in the left panel of Fig.~\ref{fig:ProdRates}. The contributions arising from different parent hadrons are shown in different colors. The shaded bands correspond to the hadron production uncertainty as defined in Sec.~\ref{sec:production}. For heavy charm and beauty hadrons, this was obtained using scale uncertainties, while for pions and kaons this corresponds to the spread of used generators. We can see that the production rate through different channels is roughly constant as long as the ALP mass is small compared to parent hadrons mass, and then plummets when approaching the respective mass. Overall, two-body decays of $B$-mesons are the most prominent production channel, with kaon decay being of similar significance for ALPs below 200~MeV. While D-meson decays provide a subdominant but still sizable contribution for ALP masses below 1~GeV, pion decays are generally of limited relevance. \medskip

\noindent \textbf{Lifetime and Decays:} In the considered mass range of $1~\mev - 10~\gev$, the only kinematically accessible ALP decay channels are $a \to ee$ and $a \to \gamma \gamma$. Following Ref.~\cite{FASER:2018eoc}, the corresponding partial decay widths are 
\be
 \Gamma_{a \to ee} &= \frac{g_{ee}^2 m_a}{8 \pi} \sqrt{1\!-\!\frac{4m_e^2}{m_a^2}} \,  \\
 \Gamma_{a \to \gamma \gamma}&=\frac{\alpha^2g^2_{ee}m_a^3}{64\pi^3m_e^2}.
 \label{eq:Lifetime}
\ee
Due to the different mass dependence, the decay into electrons dominates at low masses $m_a \lesssim 0.6~\gev$, while decays in photons will dominate at higher masses. Looking at Fig.~\ref{fig:results}, we note that ALPs in the sensitivity region of FASER will predominantly decay into electron pairs. Decay into photons will only become relevant at FASER2. The total decay width is given by their sum, and the lifetime by the inverse of the total decay width. \medskip

\noindent \textbf{Event Rate:} To obtain the event rate, the flux needs to be convoluted with the decay-in-volume probability. In the upper right panel of Fig.\ref{fig:ProdRates} we show the resulting energy spectrum of ALPs decaying in FASER, including flux uncertainties, for three benchmarks. We can see that for smaller couplings, and correspondingly longer lifetimes, the energy spectrum is broad. In contrast, for large coupling, and thus shorter lifetime, only the most energetic ALPs are able to reach and decay in FASER. Shown in the lower right panel is the corresponding ratio of the central prediction and the uncertainty. The uncertainty ratio shows no strong dependence on energy. 

\section{ALP Production Rate in B-Decays} 
\label{sec:app_b}

ALP production in beauty hadron decays is, at a partonic level, related to the flavor changing transition $b \to s \,  a$. The associated low-energy effective interaction is  $\mathcal{L}_{asb}=g_{asb} a \bar{s}_L b_R + \text{h.c.}$ and arises through loops diagrams involving the weak interactions. Following Ref.~\cite{Altmannshofer:2022izm}, the corresponding coupling is approximately given by
\be
 g_{asb} =  - \frac{g_{ee}}{m_l} \frac{3 m_W^2 m_b m_t^2 V_{ts}^* V_{tb} }{128\pi^4v^4} \, f\!\left(\frac{m_t^2}{m_W^2}\right) 
\ee
which numerically reduced to $g_{asb} = 1.1 \cdot 10^{-3} \, g_{ee}$. 

For long-lived particle searches at a far detector, we are interested in the inclusive decay branching fraction for $B \to X_s \,  a$, where $X_s$ could be any hadronic final state containing a strange quark. In the literature, there are different approaches to obtain this: 
\begin{description}
\item[Spectator Model.] Ref.~\cite{Winkler:2018qyg} uses the spectator model, which predicts a branching fraction 
\begin{equation}
\text{BR}_{B \rightarrow X_s a} = \frac{1}{\Gamma_B}  \frac{(m_B^2-m_a^2)^2}{32\pi\,m_B^3}\, |g_{asb}|^2 \, .
\end{equation}
This description follows the mass dependence of the $b \to s \, a$ decay. It is not expected to be valid close to the kinematic end-point and should only be used for $m_a < m_B - m_K$. 
\item[Exclusive Model.] Ref.~\cite{Boiarska:2019jym} estimates the inclusive branching fraction as a sum over the exclusive branching fractions for decays of the type $B \to K_i \, a$. Here $K_i$ includes  various pseudo-scalar $K$, scalar $K_0^*$, vectors $K^*$, axial-vectors $K_1$ and tensor $K_2^*$ meson states. This model is expected to underestimate the inclusive decay width at lower masses, where additional channels would need to be taken into account.
\item[Rescaled Model.] Ref.~\cite{Aloni:2018vki} suggest to approximate the inclusive decay width by $\text{BR}_{B \to X_s a} \approx 5 \times (\text{BR}_{B \to Ka}+\text{BR}_{B \to K^*a})$. Here the factor has been obtained as the ratio of branching fractions for $b \to s \mu\mu$ and $B \to K^{(*)} \mu\mu$. One should note, however, that this factor does not need to be the same for $b \rightarrow s a$ and is not expected to capture the right mass dependence. 
\end{description}
%

\begin{figure}[t]
    \includegraphics[width=0.49
    \textwidth]{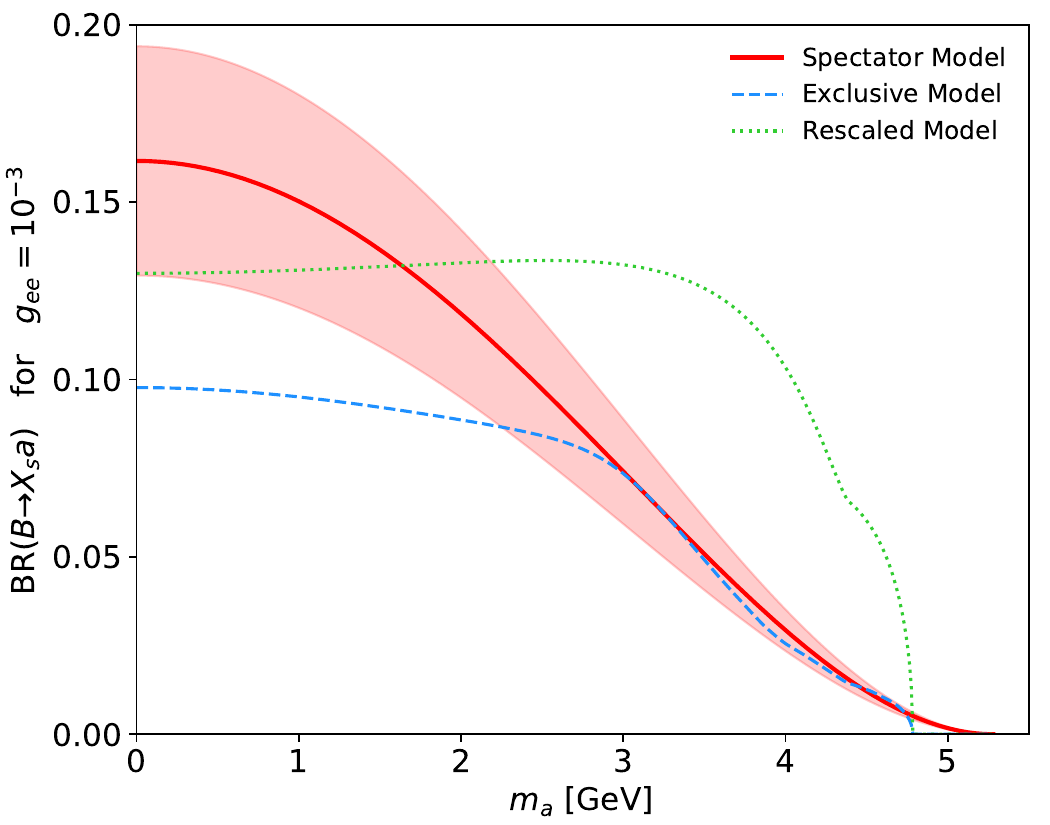}
    \caption{Comparison of BR$(B \to X_s \, a)$ predictions among the three modeling approaches for the electrophilic ALP model.}
    \label{fig:comparison}
\end{figure}

In Fig.~\ref{fig:comparison}, we present a comparison of predictions for BR$(B \to X_s \, a)$. The exclusive model closely matches the spectator model for high masses but deviates at low masses. Such a deficit is expected due to the omission of heavier kaon resonances and non-resonant kaon plus pion modes, a known effect in rare B-decays $B \to X_s \ell^+ \ell^-$~\cite{Ahmady:1995ua, BaBar:2013qry}. Conversely, the spectator model may overestimate rates at low $m_a$ by neglecting phase space suppression from the finite $X_s$ mass.  The rescaled model significantly differs from the others. It's worth noting that B-physics experiments often employ an inclusive spectator-model-like approach for studying $B \to X_s \ell^+ \ell^-$ and $B \to X_s \gamma$ decays, finding good agreement between theory predictions~\cite{Kagan:1998ym} and measurements within uncertainties~\cite{BaBar:2013qry, Belle:2014nmp}.

Regarding uncertainties, we incorporate a 20\% uncertainty factor when using the spectator model to account for the mass effects associated with the hadronic $X_s$ final state. For massive $X_s$, the phase space factor $(1-m_a^2/m_B^2)^2$ should be replaced by $\lambda_{B,X_s,a}$. For $m_{X_s}=1.5~\gev$, this results in approximately a 20\% suppression at low masses, motivating our choice.

\section{Neutrinos from B-Decays} 
\label{sec:app_fromB}

\begin{figure*}[t]
    \includegraphics[width=0.49
    \textwidth]{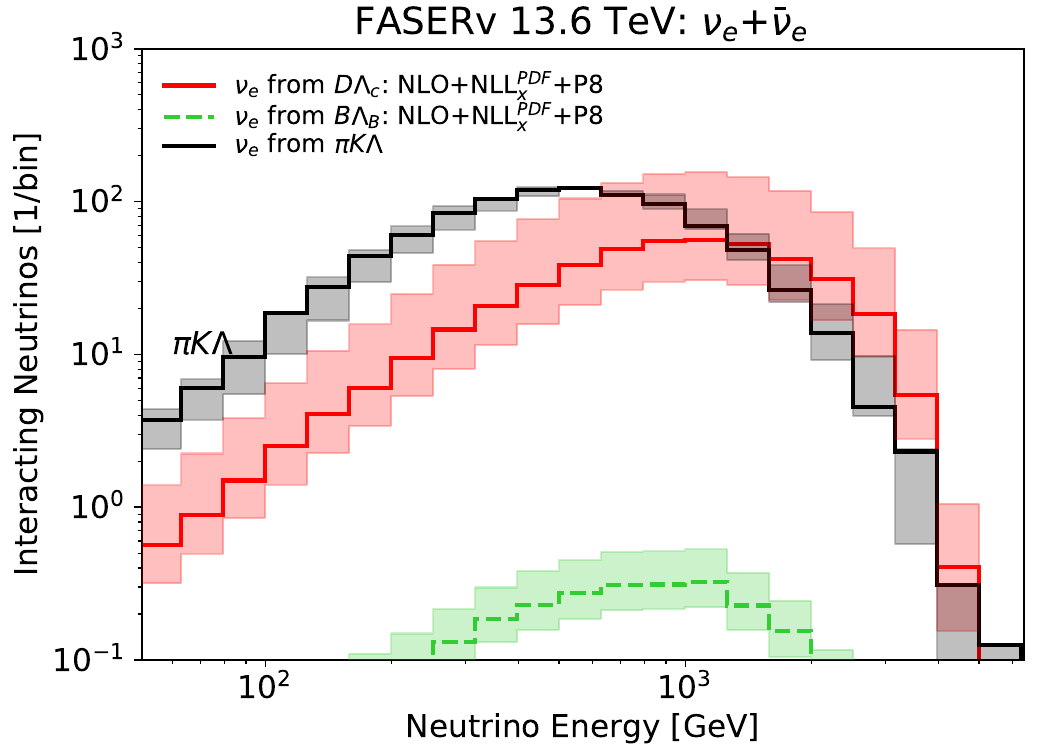}
    \includegraphics[width=0.49
    \textwidth]{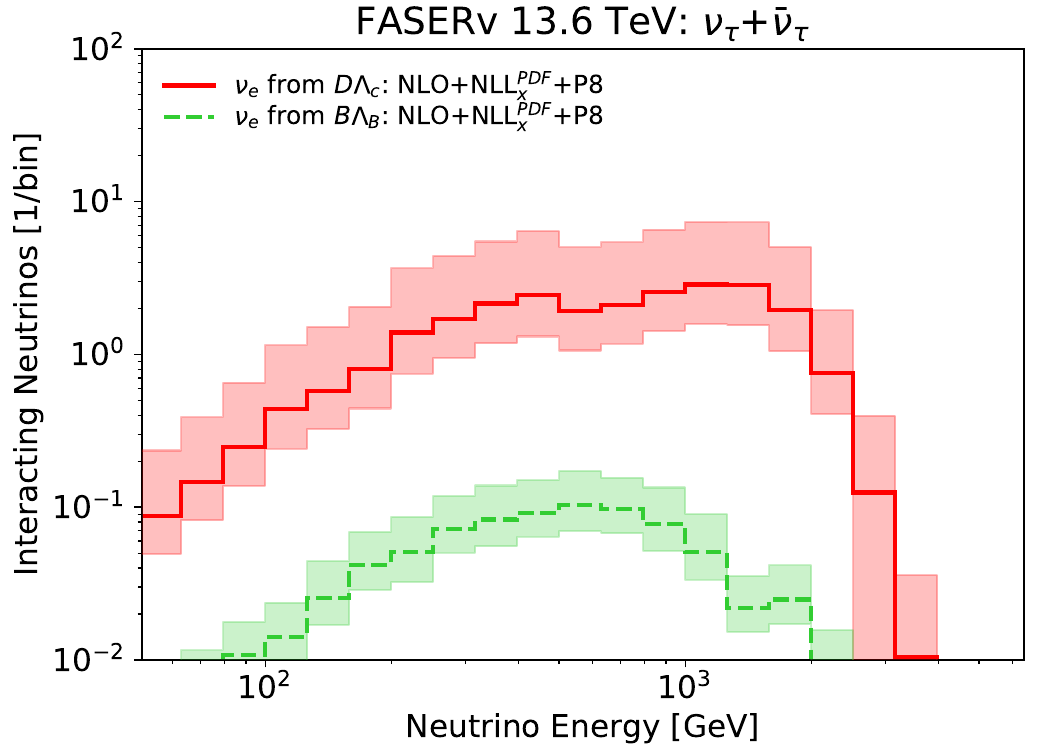}
    \caption{Predicted energy spectrum of electron neutrinos (left) and tau neutrinos (right) from light, charm and beauty hadrons decays at FASER$\nu$. We show the central prediction as solid line and the associated uncertainty as light shaded band. }
    \label{fig:nu_withB}
\end{figure*}

In the main part of the text, we have presented results for the dominant neutrino flux components from light and charm hadron decays. Here, for completeness, we also show results for the subleading neutrino flux component from beauty hadrons decays. The energy spectrum of electron and tau neutrinos originating from beauty hadron decay and interacting in FASER$\nu$ is shown in Fig.~\ref{fig:nu_withB}. We can see that the neutrino flux from beauty hadrons contributes less than 0.1\% to the number of electron neutrino interactions, but about 3\% to the number of tau neutrino interactions. These results qualitatively agree with those obtained at leading-order using \textsc{Pythia~8.2} presented in Ref.~\cite{FASER:2019dxq}. We predict  $2.54^{+ 1.66}_{ - 0.8} \ \nu_e$, $ 2.52^{ + 1.65}_{ - 0.82} \ \nu_\mu$ and $0.79^{ + 0.53}_{ - 0.25 }\ \nu_\tau$  charged current neutrino interactions from bottom hadron decays to occur in FASER$\nu$ during Run~3 with $200~\ifb$.

\bibliographystyle{spphys}
\bibliography{references}

\end{document}